\documentclass[floatfix, preprintnumbers, showpacs, twocolumn, amsmath, amssymb, aps]{revtex4}

\usepackage{bm}
\usepackage{amsmath}
\usepackage{amsthm}
\usepackage{amssymb}
\usepackage{graphicx}
\usepackage{bbold}
\usepackage{url}
\usepackage[caption=false]{subfig}

\newcommand{\ket}[1]{\left| #1 \right>}

\begin{document}
\title{Geometries for universal quantum computation with matchgates}

\author{Daniel J. Brod}
\email{brod@if.uff.br}
\affiliation{Instituto de F\'isica, Universidade Federal Fluminense, Av. Gal. Milton Tavares de Souza s/n, 
Gragoat\'a, Niter\'oi, RJ, 24210-340, Brazil}

\author{Ernesto F. Galv\~ao}
\email{ernesto@if.uff.br}
\affiliation{Instituto de F\'isica, Universidade Federal Fluminense, Av. Gal. Milton Tavares de Souza s/n, 
Gragoat\'a, Niter\'oi, RJ, 24210-340, Brazil}

\date{\today}

\begin{abstract}
Matchgates are a group of two-qubit gates associated with free fermions. They are classically simulatable if restricted to act between nearest neighbors on a one-dimensional chain, but become universal for quantum computation with longer-range interactions. We describe various alternative geometries with nearest-neighbor interactions that result in universal quantum computation with matchgates only, including subtle departures from the chain. Our results pave the way for new quantum computer architectures that rely solely on the simple interactions associated with matchgates. 
\end{abstract}
\pacs{03.67.Lx, 05.30.Fk}
\maketitle

\section{Introduction}
There is an on-going research effort to identify the quantum features responsible for the computational advantage of quantum over classical computers. One way of doing that is to propose restricted quantum scenarios which can be efficiently simulated on a classical computer, and then add further resources that provably enable universal quantum computation. This classical-to-quantum transition in computational power happens, for example, when we add entangling gates to the set of all single-qubit unitaries or when we add a generic gate to the Clifford group of unitaries. Another motivation for this approach is the possibility of identifying alternative implementations of quantum computers, which may be more practical, versatile or especially suited to a given physical system.

Here we investigate the transition from classical to quantum computational power in the context of matchgates, a group of two-qubit gates that were proposed in graph-theoretical terms by Valiant in 2002 \cite{Valiant02}. Matchgates acting only on nearest-neighbors on a 1D chain of qubits are classically simulatable, a fact that was traced back to their equivalence to a system of non-interacting fermions via the Jordan-Wigner transformation \cite{Terhal02}.  Interestingly, they allow for universal quantum computation if next-nearest neighbor matchgate interactions are possible, or by the equivalent use of the SWAP gate \cite{Jozsa08b} or other non-matchgate unitaries \cite{Brod11}.

In this paper we ask what happens when matchgates act on nearest-neighboring qubits arranged in geometries other than the linear chain. We prove that there are many possible interaction geometries which enable matchgates to do universal quantum computation, ranging from near-trivial modifications of the chain to quite different graph families, such as star graphs and binary trees. This also partially answers the question of whether the classical simulability of nearest-neighbor matchgates holds beyond the linear chain, an open question up until now (due, in part, to the fact that the Jordan-Wigner transformation cannot be trivially generalized to nonlinear geometries). 

Matchgates arise naturally from interactions in various physical systems, especially in the form of the much studied anisotropic Heisenberg (or XY) interaction \cite{Lidar01, Wu02b, Kempe02}, which indicates that the geometrical arrangements we propose may be suitable for practical implementations of quantum computers.

\section{Quantum computation with matchgates}
Let $G(A,B)$ denote the 2-qubit gate that acts as unitaries $A$ and $B$, respectively, on the even and odd parity subspaces of the 2-qubit Hilbert space:
\begin{equation} \label{Matchgate}
G(A,B) = \left(\begin{array}{cccc}
A_{11} & 0 & 0 & A_{12} \\
0 & B_{11} & B_{12} & 0 \\
0 & B_{21} & B_{22} & 0 \\
A_{21} & 0 & 0 & A_{22}
\end{array}\right).
\end{equation}
If it satisfies the additional constraint that det$A$ = det$B$, $G(A,B)$ is a \textit{matchgate}. 

Valiant introduced matchgates in graph-theoretical terms and provided an explicit algorithm for efficiently simulating matchgates between nearest-neighboring qubits on a linear chain \cite{Valiant02}. Curiously, this theoretical construction was later shown to be exactly equivalent to the simulation of non-interacting fermions; one problem is mapped onto the other via the Jordan-Wigner transformation \cite{Terhal02}. In 2008 Jozsa and Miyake complemented this simulability result with the missing element for universal quantum computation: the SWAP gate, or alternatively the possibility of acting with matchgates between next-nearest-neighbors on the linear chain \cite{Jozsa08b}. While their main result required encoding each logical qubit into 4 physical qubits, a more economical 2-to-1 encoding was also presented, but which required matchgate interactions between third neighbors. 
\begin{figure}
\centering
\includegraphics[width=0.32\textwidth]{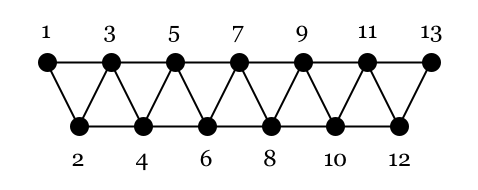}  
\caption{In the ``triangular ladder'' array, qubits have a one-to-one correspondence to qubits in a linear array such that nearest neighbors on the former correspond to nearest and next-nearest neighbors on the latter.}	
\label{triangladder}
\end{figure}

These results can alternatively be stated in terms of the geometrical arrangement of the qubits. Remark that nearest- and next-nearest neighbors of the 1D chain can be mapped into nearest-neighbors in the ``triangular ladder" graph of Fig. \ref{triangladder}. Thus, as was already implicit in previous works \cite{Kempe02}, the computational power of nearest-neighbor matchgates jumps from (sub-)classical \cite{Nest10} to quantum universal by rearranging the qubits of the linear chain to form a triangular ladder. In what follows we identify many other graph families that also allow for universal computation with matchgates acting only between nearest-neighbors, including much simpler alternatives to the triangular ladder. We need first to develop a couple of tools.

\section{The ``hair comb'' graph}
The SWAP gate [$= G(I,X)$] is parity-preserving but is not a matchgate, as it does not satisfy the determinant condition. This is why the connectivity restrictions are so important -- quantum information is not free to move around if all we have is matchgates.  There is, however, a matchgate which is similar to the SWAP, which is the fermionic-SWAP (f-SWAP): 
\begin{equation} \label{fSwap}
\textrm{f-SWAP} = G(Z,X) = \left(\begin{array}{cccc}
1 & 0 & 0 & 0 \\
0 & 0 & 1 & 0 \\
0 & 1 & 0 & 0 \\
0 & 0 & 0 & -1
\end{array}\right).
\end{equation}
It cannot be used to swap the states of two arbitrary qubits, as the minus sign induced in the $\ket{11}$ state can have an entangling effect (the f-SWAP is actually a perfect entangler \cite{Brod11}). If, however, we know for certain that one of the input qubits is in the $\ket{0}$ state, then the f-SWAP acts exactly as the SWAP, a trick which was already exploited in \cite{Jozsa08b} with this exact purpose. Given the simulability of nearest-neighbor matchgates on the 1D chain, this f-SWAP trick clearly cannot be used to replace next-nearest neighbor interactions. However, if we modify the geometry by introducing a T structure on the chain (see Fig. \ref{FigT}-a), we can initialize the appended ancilla in the $\ket{0}$ state and use it to circumvent the need for next-nearest neighbor interactions. To implement any matchgate $G$ between qubits $i-1$ and $i+1$ of Fig. \ref{FigT}-a, we can apply the sequence:
\begin{figure}[t]
\centering
\subfloat[f-SWAP gadget]{\includegraphics[width=0.2\textwidth]{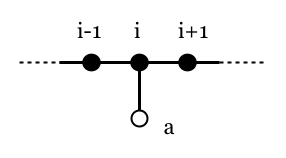}}
\subfloat[H-gadget]{\includegraphics[width=0.2\textwidth]{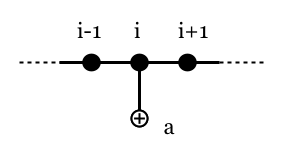}} \\
\caption{The T structure. (a) The f-SWAP gadget simulates a next-nearest neighbor interaction between qubits $i-1$ and $i+1$ using f-SWAP gates and a $\ket{0}$ ancilla. (b) The H-gadget implements an $H$ gate on qubit $i$ using a matchgate $G(H,H)$ between $i$ and a $\ket{+}$ ancilla.}
\label{FigT}
\end{figure} 
\begin{equation} \label{Swapgadget}
I_{i} \otimes G_{[i-1,i+1]} \otimes I_{a} = F_{[a,i]} F_{[i,i-1]} G_{[i,i+1]} F_{[i,i-1]} F_{[a,i]},
\end{equation}
where $F$ is a shorthand for the f-SWAP and subscripts represent the pair of qubits on which the gate acts, as labeled in Fig. \ref{FigT}-a. We see that the f-SWAP and $\ket{0}$ ancilla effectively simulate a next-nearest neighbor interaction by temporarily ``hiding'' the information of qubit $i$ in the appended qubit. We dub this use of the f-SWAP trick on the T structure a ``f-SWAP gadget".
\begin{figure}[t]
\centering
\includegraphics[width=0.3\textwidth]{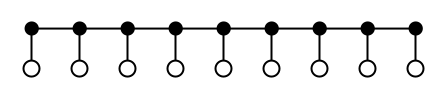}
\caption{Hair comb graph. By repeating the f-SWAP-gadget throughout the line we can simulate interactions between any pair of next-nearest neighbors, enabling universal computation.}
\label{HairComb}
\end{figure} 

If the states of all qubits can be selectively ``hidden" in ancillas, all next-nearest neighbor interactions become possible and we achieve universal quantum computation with matchgates only, by the result of Jozsa and Miyake \cite{Jozsa08b}. This can be done by introducing multiple T structures on the chain, resulting in the ``hair comb'' graph of Fig. \ref{HairComb}. Furthermore, a trivial extension of the sequence of Eq. (\ref{Swapgadget}) enables us to interact qubits which were originally third neighbors, which means we can use the more economical 2-to-1 encoding of \cite{Jozsa08b}. This is our first result of a graph for universal quantum computation with matchgates. A quick resource count shows that the protocol described enables universal computation using $4$ additional f-SWAPs for each SWAP and the same number of qubits as the encoded quantum circuit of \cite{Jozsa08b}, in a simpler graph than the triangular ladder of Fig. \ref{triangladder}.

An interesting alternative demonstration of this result can be given as follows. A standard universal set for quantum computing consists of single-qubit $Z$-rotations, the Hadamard ($H$) gate and any two-qubit entangling gate \cite{LivroNielsen}. As noted in \cite{Brod11}, matchgates include all single-qubit $Z$-rotations and many perfect entanglers, which means that matchgates become universal when supplemented by $H$. We now propose a scheme to implement the Hadamard $H$ gate with matchgates only, based on a simple observation:  
\begin{equation} \label{HGadget}
G(H,H) (\ket{\psi} \ket{+}) = (H \ket{\psi}) \ket{+}.
\end{equation}
Since $G(H,H)$ is a matchgate, this means that initializing ancilla qubits in the $\ket{+}$ state can replace the need for $H$ gates. This clearly cannot be used for universal computation on the 1D chain, as in this geometry we cannot arrange the required $\ket{+}$ ancillas so as to apply $H$ wherever necessary.  However, as shown in Fig. \ref{FigT}-b, we can again modify the geometry and introduce T structures where appended $\ket{+}$ ancillas effectively apply a $H$ gate using the matchgate sequence of Eq. (\ref{HGadget}). We dub this use of the $G(H,H)$ gate on the T structure an ``H-gadget'', see Fig. \ref{FigT}-b. By repeating this structure at every chain qubit we can implement the $H$ gate on any of them, again enabling universal quantum computation with nearest-neighbor matchgates on the hair comb graph of Fig. \ref{HairComb}. Resource-wise, this scheme requires as many gates as the original quantum circuit and twice as many qubits, and does not require encoding as the f-SWAP gadget scheme does.
\begin{figure}[t]
\subfloat[Square lattice]{\includegraphics[width=0.20\textwidth]{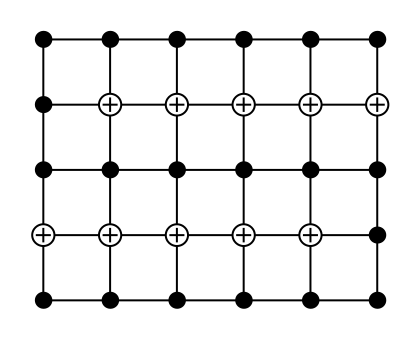}}
\subfloat[Cycle with appended vertex]{\includegraphics[width=0.20\textwidth]{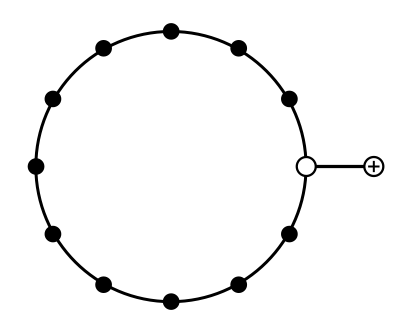}} \\
\subfloat[An example of $3$-regular graph]{\includegraphics[width=0.20\textwidth]{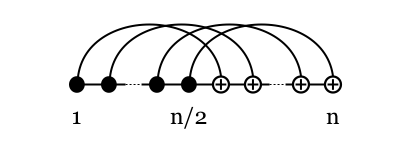}}
\subfloat[Chain with appended vertex]{\includegraphics[width=0.20\textwidth]{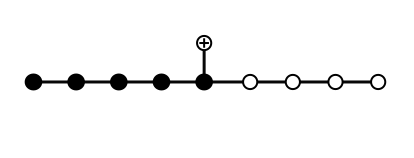}} \\
\subfloat[Star graph]{\includegraphics[width=0.20\textwidth]{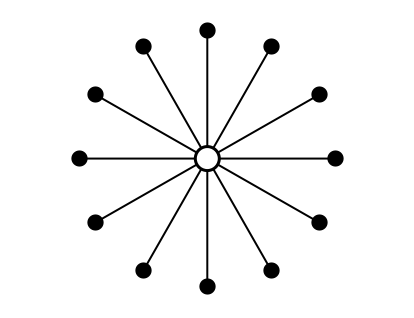}}
\subfloat[Wheel graph]{\includegraphics[width=0.20\textwidth]{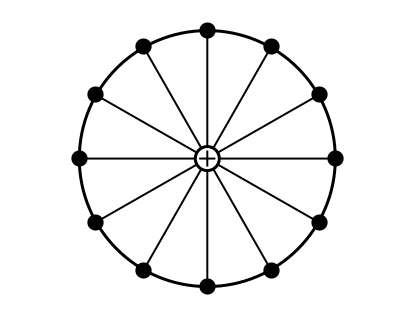}} \\\caption{Several geometries for nearest-neighbor matchgates interactions which are universal for quantum computation. White circles and circles with crosses represent one possible placement of the $\ket{0}$ and $\ket{+}$ ancillas, respectively, which makes the universality of the graphs explicit by the arguments in the main text.}
\label{Examples}
\end{figure} 

We can extend these results by noticing that, in both cases, the ancilla's state is left unaltered by the gadgets [cf. Eqs. (\ref{Swapgadget}) and (\ref{HGadget})]. Physically, this means that some computational qubits can share the same ancilla, going as far as the wheel graph, where a single ancilla is connected to every qubit on the cycle (see Fig. \ref{Examples}-f). Note that the ancillas used in H-gadgets never get entangled with other qubits, and so can be prepared, used and discarded on the fly if necessary.

\section{General graphs}
As we have seen, the ``hair comb'' and wheel graphs allow efficient schemes for universal computation using matchgates only. A large number of graphs which arise naturally in physical systems can be trivially shown to contain those graphs as subgraphs, such as all 2D regular lattices common in solid state physics, and all demi-, semi- and regular tessellations \cite{tessellations}. 

However, as we show now, there are strikingly simpler graphs which are also universal for quantum computation if we allow for a larger (but still polynomial) overhead in resources. Consider first a modification of the cycle graph of $n$ vertices $C_n$, where a single additional vertex is appended to any chosen vertex $i$ of the cycle (see Fig. \ref{Examples}-b). Locally, this appended vertex is equivalent to the T structure of Fig. \ref{FigT}, which means we can use the H-gadget on it, as before. Notice also that if we assign one qubit of the cycle as a $\ket{0}$ ancilla, we can use the f-SWAP trick to move it freely around the cycle. Each complete clockwise cycle of the ancilla shifts the state of all other qubits by one step in the counter-clockwise direction (and vice-versa). By f-SWAPPing the ancilla around a sufficient number of times, we can place any qubit on position $i$, where it can undergo an $H$ gate via our H-gadget. As before, this allows for universal computation, since the overhead in resources amounts to 2 additional qubits and at most $n^2/2$ f-SWAP gates for every $H$ gate in the original circuit. This also proves the universality of any graph family with a sufficiently large cycle (i.e. one that grows polynomially with the size of the graph), with the obvious exception of the cycle graph itself. 

Together with the previous results concerning the T structure, this cycle result provides a powerful toolbox for demonstrating the universality of more general families of graphs. For instance, it is easy to show that any graph which contains a polynomially long path where every vertex has degree greater than $2$ is universal. This becomes clear by recognizing that every vertex in this long path is either part of a cycle or of a T structure, and our previous results can be combined to provide a universal scheme. Notice that these tools may not be sufficient to decide whether an arbitrary graph is universal, as it is NP-hard to find the longest path or cycle in a general graph \cite{Garey79}. Nevertheless, they still contemplate any graph family which is known to have a polynomially long path.

Let us now consider a very slight deviation from the linear chain: we connect a single extra vertex to the central chain vertex, as in Fig. \ref{Examples}-d. We can assign the first half of the vertices of the chain as computational, the rest of the chain as $\ket{0}$ states and the appended vertex as a $\ket{+}$ state so as to apply the H-gadget on the central qubit. By using the f-SWAP trick we can move the state of the computational qubits back and forth, placing any chosen qubit state at the H-gadget site. This results in universal quantum computation with an overhead of at most $n^2$ f-SWAPs per $H$ gate. The scheme, as described, works as long as the extra vertex is sufficiently far away from the endpoints of the chain. If the vertex is logarithmically close to one of the endpoints, the number of qubit states that can be placed at the H-gadget site is logarithmic, and thus insufficient. However, in the Appendix we show an alternative universal scheme that works with the extra vertex placed at any point (with the natural exception of the endpoints), at the cost of using a 2-to-1 encoding for the logical qubits. 

It is surprising that such small deviations from the linear chain already result in universal quantum computation. Starting with the chain, it is sufficient to either add a single edge (creating a cycle) or append a single ancilla qubit - which can even be just one vertex away from the endpoint. This is similar in spirit to a result in the context of quantum control theory \cite{Burgarth10}, where it is shown that, if all neighboring qubits in the linear chain interact with each other via a specific background (hence uncontrollable) two-qubit interaction, it suffices to have full control of only the two first qubits of the chain. Fast, arbitrary gates on the first two qubits are enough to control the state of the whole chain, driving it to do universal computation.

All graph families we have proven so far to be universal for matchgate quantum computation use a long path as a ``backbone'' for the computation, and consist of progressively small deviations from the linear chain. Our tools can also be used to prove similar results for a number of graph families which differ significantly from the chain. Take the complete binary tree \cite{binarytree} with $n$ leaves, for which the longest path is of size $\log{n}$. By assigning the leaves as computational qubits and filling the branches with $\ket{0}$ ancillas, we can use f-SWAP to move the states of any pair of computational qubits together, interact them and return to the original configuration. This scheme implements each computational matchgate at a cost of a logarithmic number of f-SWAP gates. It is clear that any graph family where a polynomial-sized set of vertices can be connected to all other vertices in the set by a path of $\ket{0}$ ancilla qubits will be universal for matchgate quantum computation. An interesting, extremal example is the star graph (see Fig. \ref{Examples}-e), where the only edges are those between a central vertex and all others.

Fig. \ref{Examples} displays various qualitatively different examples of graph families we have proven to be universal for quantum computation using nearest-neighbor matchgates only. In each of these graphs we mark some vertices as $\ket{0}$ or $\ket{+}$ ancillas to highlight one possible universal scheme. For what follows it is important to point out that, while we used the H-gadget in several cases in the interest of displaying economical protocols, it is not strictly necessary in order to demonstrate universality in any situation we have considered. The f-SWAP gadget can replace the H-gadget on any geometry with only a polynomial overhead in space and time due respectively to the use of encoding and additional f-SWAP gates.

\section{Physical implementations}
Our results will be experimentally relevant in any situation where matchgate interactions arise naturally. As is well-known, the group of matchgates is generated by the set of two-qubit Hamiltonians: $\{ XX, YY, XY, YX, IZ, ZI \}$ \cite{Terhal02}.  As a practical example, let us consider the much-studied anisotropic Heisenberg Hamiltonian $XX+YY$ \cite{Kempe02, Lidar01}. It was proven in \cite{Kempe02} that the matchgates generated by this Hamiltonian are universal for quantum computation when allowed to act on next-nearest neighbors on the chain, or equivalently between nearest-neighbors on the triangular ladder of Fig. \ref{triangladder}. 

Now let us write the f-SWAP gate in a form that explicits the underlying interactions:
\begin{equation*}
\textrm{f-SWAP} = \textrm{Exp} \left [ i \frac{\pi}{4} (IZ + ZI) \right] \textrm{Exp} \left [ i \frac{\pi}{4} (XX + YY) \right].
\end{equation*}
We see that besides the $XX+YY$ interaction, the f-SWAP gate also requires single-qubit $Z$ rotations. This is a possibility in a number of quantum computer implementations, such as cavity QED with atoms \cite{Zheng00} or quantum dots \cite{Imamoglu99}, and electrons in liquid Helium \cite{Mozyrsky01}. Given the result of \cite{Kempe02}, our work suggests multiple choices of interaction geometries that may simplify the design of universal quantum computers using these systems.

We have also described an alternative matchgate quantum computing scheme that uses H-gadgets instead of the f-SWAP trick. One advantage of this alternative is that the required $\ket{+}$ ancillas never get entangled with the other qubits, and so can be replaced at will between uses or may even consist of a single, recyclable moving ancillary qubit. We note that the $G(H,H)$ gate cannot be decomposed in terms of the same interactions as the f-SWAP; it would be interesting to identify a physical implementation where it occurs naturally. 

\section{Open questions}
In summary, we have proven that nearest-neighbor matchgates are universal for quantum computation in very diverse geometrical settings: on a graph with a large enough cycle, on the star graph, and on a chain with a single appended vertex, for example. The difference between this last case and the simulatable linear chain is minimal, as the single extra qubit may remain in the $\ket{+}$ state during the computation, never entangling with the chain qubits. 

As a consequence, now we know that it is not possible to efficiently simulate matchgates between nearest-neighbors in all these graph families. It would be interesting if one could find graph families where matchgates are neither simulatable nor universal for quantum computation, an intermediate case of computing power that has drawn much recent interest \cite{Bremner10, Aaronson11}. 

As we have seen, useful sets of matchgates can be generated with the $XX+YY$ Hamiltonian supplemented with single-qubit $Z$ rotations, present  naturally in systems such as cavity QED with quantum dots and atoms, and electrons in liquid Helium. Our results then suggest different architectures that could enable universal quantum computation in these systems. We leave as an open question whether our tools can be adapted to obtain new architectures where only a single interaction is used (as in \cite{Kempe02} and \cite{DiVincenzo00}).

Finally we point out that, by applying the Jordan-Wigner transformation on the 1D chain, one can show that nearest-neighbor matchgates correspond to noninteracting fermions, while matchgates between more distant qubits translate into fermionic interactions \cite{Jozsa08b}. As our results show the need for only very small deviations from the linear chain, it would be interesting to investigate the implications for proposals for quantum computation with fermions \cite{Bravyi02, Terhal02, Beenakker04}.

\begin{acknowledgments}
We are grateful to Daniel Burgarth for helpful discussions. We acknowledge financial support by the Instituto Nacional de Ci\^{e}ncia e Tecnologia de Informa\c{c}\~{a}o Qu\^{a}ntica (INCT-IQ/CNPq - Brazil).
\end{acknowledgments}

\appendix*
\section{}

In this appendix we show how to use codification and the f-SWAP gate to achieve universal computation with matchgates on the chain with an appended vertex of Fig. \ref{Examples}-d, even if the appended vertex is placed near the chain endpoints. For this, we need to encode each logical qubit into the even parity subspace of two physical qubits as done in \cite{Jozsa08b, Brod11}:
\begin{align} 
\ket{0}_L & = \ket{00} \notag \\
\ket{1}_L & = \ket{11},
\end{align}
where subscript $L$ denotes a logical qubit. Let $\ket{\psi}$ and $\ket{\phi}_L$ represent the arbitrary states of a physical and a logical qubit, respectively. Then, it is easy to check that 
\begin{equation} \label{logicalSWAP}
F_{[1,2]} F_{[2,3]} \ket{\psi} \ket{\phi}_L = \ket{\phi}_L \ket{\psi},
\end{equation}
where $F_{[i,j]}$ denotes the f-SWAP acting on qubits $i$ and $j$. That is, consecutive applications of the f-SWAP can move the state of an arbitrary qubit ``through'' a logical qubit. This only works if a complete logical qubit is traversed at a time. The reason behind this is that the f-SWAP gate will either act as the SWAP (if the logical qubit is in the $\ket{00}$ state) or, if it induces a minus sign, it does so twice (if the logical qubit is in the $\ket{11}$ state). Thus, the f-SWAP can be used to move blocks of two physical qubits around, as long as they comprise the same logical qubit.

Now consider the universal set constructed in \cite{Jozsa08b, Brod11}. A logical single-qubit gate $A$ is implemented by the matchgate $G(A,A)$ acting on the two physical qubits that form a logical qubit. A logical entangling gate can be implemented by the sequence:
\begin{equation} \label{controlZ}
CZ_{23} = G(H,H)G(X,X) \; \textrm{SWAP} \; G(H,H),
\end{equation}
where the first logical qubit is encoded in physical qubits $1$ and $2$, and the second logical qubit is encoded in physical qubits $3$ and $4$. All gates in the above equation are between qubits $2$ and $3$, and this $CZ_{[2,3]}$ gate induces an effective $CZ$ gate between the encoded qubits. This cannot be implemented in the linear chain with only matchgates because of the SWAP gate in Eq. (\ref{controlZ}). However, as mentioned in Section IV of \cite{Brod11}, this SWAP gate can be decomposed in terms of matchgates and Hadamard gates on both qubits $2$ and $3$.

It is clear that we can adapt this result to our context using the tools developed in the main text. Suppose that the qubits are aligned on a linear chain, ordered such that every pair $(2i-1, 2i)$ comprises logical qubit $i$. Since logical single-qubit gates are easy to implement with matchgates, all that is necessary is to implement a logical $CZ$ gate between the two logical qubits at the chain's end, which involve only physical qubits $1$ to $4$. If the computation being performed requires a $CZ$ gate between another pair of logical qubits, we simply use the trick of Eq. (\ref{logicalSWAP}) to SWAP their state to the endpoint.

\begin{figure}[t]
\centering
\includegraphics[width=0.25\textwidth]{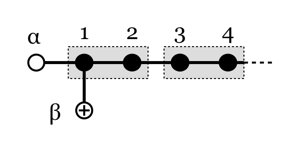}
\caption{Scheme for universal computation on the 1D chain and one extra qubit appended at the second vertex. The shaded rectangles delimit the (initial positions of) logical qubits. A combination of f-SWAP gates and the H-gadget enables us to implement a $CZ$ gate between these logical qubits (see main text for the explicit sequence of operations).}
\label{FigAppend}
\end{figure}

As shown in \cite{Jozsa08b}, the logical $CZ$ can be implemented by a physical $CZ$ between any pair consisting of one physical qubit from each logical qubit. By the previous discussion and section IV of \cite{Brod11}, this reduces to implementing the $H$ gate on both these physical qubits. To achieve that, consider the configuration of Fig. \ref{FigAppend}, where we appended two ancilla qubits, one in the $\ket{0}$ and one in the $\ket{+}$ state, to the endpoint of the chain. Assume w.l.o.g. that the logical qubits we wish to act with the $CZ$ on already occupy positions $(1,2)$ and $(3,4)$, labeled as in Fig. \ref{FigAppend}. The scheme to implement the $CZ$ proceeds as follows.

Using two consecutive f-SWAPS, swap the state of physical qubit $3$ across the first logical qubit (i.e. to position $1$). At this point, positions $1$ and $2$ are each occupied by one component of a different logical qubit. Now use the f-SWAP trick described in the main text first to swap pair $(\alpha,1)$ and then pair $(\beta,1)$. This places the $\ket{+}$ ancilla adjacent to the two qubits we wish to apply the $H$ on. Thus, by repeating this process back and forth we can alternatively implement a matchgate between these two physical qubits or a $H$ gate on each of them. Consequently, by the result of \cite{Brod11}, this effectively allows us to implement an entangling gate between the two logical qubits. Notice that, after application of the $H$ gate, the physical qubits temporarily leave the encoded space. At this point Eq. (\ref{logicalSWAP}) no longer applies, and any swapping must be made with the aid of the $\ket{0}$ ancilla, as done throughout the main text. Only at the end of the sequence of Eq. (\ref{controlZ}) do the qubits return to the encoded space and the trick of Eq. (\ref{logicalSWAP}) can be used to return the logical qubits to their original positions.

The scheme described above allows us to implement a logical entangling gate between the first two logical qubits and, by extension, between any pair of logical qubits throughout the circuit. This shows that the 1D chain with a single vertex appended to the second chain vertex is universal for quantum computation with matchgates. In fact, by the previous reasoning it is easy to see that this extra vertex can be appended absolutely anywhere, with the exception of the chain endpoints, completing the demonstration for the claim made in the main text.

\bibliographystyle{unsrt}

\end{document}